\documentstyle [12pt]{article}
\newcommand{\id}{\mbox{{\it {\bf 1}}}}
\let\uml=\" \let\acut=\'  
 \def\e{{\rm e}} \def\phi{\varphi}

 \def\a{\alpha} \def\b{\beta}
\def\g{\gamma}  \def\d{\delta}  
\def\l{\lambda}  
\def\s{\sigma} \def\tr{\mathop{\hbox{\rm tr}}\nolimits}

 \def\res{\mathop{\rm res}\limits}

 \def\dd{\partial}

\def\one#1{#1^{\raise5pt\hbox{$\scriptstyle\!\!\!\!1$}}\,{}}
\def\two#1{#1^{\raise5pt\hbox{$\scriptstyle\!\!\!\!2$}}\,{}}
\def\three#1{#1^{\raise5pt\hbox{$\scriptstyle\!\!\!\!3$}}\,{}}

 \def\comment#1{}
 
\def\endproof{\hfill\rule{2mm}{2mm}}
\def\beq{\begin{equation}} \def\eeq{\end{equation}}
\def\be{\begin{displaymath}} \def\ee{\end{displaymath}}
\def\bea{\begin{eqnarray}} \def\eea{\end{eqnarray}}
\def\beas{\begin{eqnarray*}} \def\eeas{\end{eqnarray*}}
\def\bds{\begin{description}} \def\eds{\end{description}}
\def\bmat{\left(\begin{array}} \def\emat{\end{array}\right)}

\def\Ref#1{(\ref{#1})}
\def\?{(?)\marginpar{|?}}

\renewcommand{\theequation}{\thesection.\arabic{equation}}
\newcounter{subequation}[equation]
\makeatletter
 
\expandafter\let\expandafter\reset@font\csname reset@font\endcsname
 
\def\subeqnarray{\arraycolsep1pt
    \def\@eqnnum\stepcounter##1{\stepcounter{subequation}%
        {\reset@font\rm(\theequation\alph{subequation})}}
\jot5mm     \eqnarray}

\makeatother\newcommand{\newsection}[1]{\vspace{10mm}
\pagebreak[3]\addtocounter{section}{1}\setcounter{equation}{0}
\setcounter{subsection}{0}\setcounter{footnote}{0}
 
\begin{flushleft}{\Large\bf \thesection. #1}
\end{flushleft}\nopagebreak\medskip\nopagebreak}

\newfont{\bbd}{msbm10 scaled\magstep1} \def\C{\hbox{\bbd C}}                  
\def\R{\hbox{\bbd R}}  \def\Z{\hbox{\bbd Z}}

\newfont{\frak}{eufm10 scaled\magstep1}

\newcounter{remctr}

\newcommand{\tfrac}[2]{{\textstyle\frac{#1}{#2}}}
\begin{document}
\begin{center} \LARGE\bf
Separation of variables for the Ruijsenaars system
\end{center}
\vskip 1cm
\begin{center}
V.B.~Kuznetsov$\,{}^\dagger$%
\footnote{On leave from: Department of Mathematical and Computational
Physics, Institute of Physics, St.~Petersburg University, St.~Petersburg
198904, Russian Federation. E-mail: \tt vadim@amsta. leeds.ac.uk}, 
F.W.~Nijhoff$\,{}^\dagger$%
\footnote{E-mail: \tt frank@amsta.leeds.ac.uk}
and E.K.~Sklyanin$\,{}^\ddagger$%
\footnote{On leave from: Steklov Mathematical Institute,
Fontanka 27, St.~Petersburg 191011, Russian Federation.
E-mails: {\tt sklyanin@kurims.kyoto-u.ac.jp} and 
{\tt sklyanin@pdmi.ras.ru}} 
\vskip 0.7cm
${}^\dagger\;$Department of Applied Mathematical Studies,\\  
University of Leeds, Leeds LS2 9JT, UK
\vskip 0.4cm
${}^\ddagger\;$Research Institute for Mathematical Sciences, \\
Kyoto University, Kyoto 606, Japan
\end{center}
\vskip 2cm
\begin{center}
\bf Abstract
\end{center}
We construct a separation of variables for the classical 
$n$-particle Ruijsenaars system (the relativistic analog of the elliptic
Calogero-Moser
system). The separated coordinates appear as the poles of the properly 
normalised eigenvector (Baker-Akhiezer function) of the corresponding Lax 
matrix. Two different normalisations of the BA functions are analysed.
The canonicity of the separated variables is verified with the use of 
$r$-matrix technique. The explicit expressions for the generating function
of the separating canonical transform are given in the simplest cases
$n=2$ and $n=3$. Taking nonrelativistic limit we also construct
a separation of variables for the elliptic Calogero-Moser system. 
\vskip 3cm

{\tt solv-int/9701004}
\vskip 2cm
\newpage
%
%%%%%%%%%%%%%%%%%%%%%%%%%%%%%%  SECTION 1  %%%%%%%%%%%%%%%%%%%%%%%%%%%%%%%%%%%
%
\newsection{Introduction}
\setcounter{equation}{0}
One of the most powerful methods in studies of Liouville 
integrable systems is that of {\it Separation of Variables} (SoV). 
Originated with the development of the Hamiltonian mechanics as
a method to solve the Hamilton-Jacobi equation for particular
Hamiltonians, nowadays it has been applied to many families 
of finite-dimensional (Liouville) integrable systems (see recent
review \cite{S1}). 

{}For a very long time a great deal of attention has been 
given to so-called coordinate separation of variables
or to separation in the configuration space
(see, for instance, \cite{Ka,S4,K1,K2,K3,kkm94,kkm95,S1} 
and references therein).
In this case the separation variables $u_j$ do not depend on the momenta
$p_i$ and are functions of the coordinates $x_i$ only:
\be
u_j=u_j(x_1,\ldots,x_N)\,.
\ee
Such kinds of integrable systems admitting a coordinate 
(local) separation of variables were studied in detail,
although in the same time it was understood that far not
every Liouville integrable system can be separated through such a transition
to new ``coordinates'' $u_j$. The class of admissible
transformations should be enlarged for a generic integrable system
up to a general canonical transformation
\be
u_j=u_j(x_1,\ldots,x_N,p_1,\ldots,p_N)\,,\qquad 
v_j=v_j(x_1,\ldots,x_N,p_1,\ldots,p_N)\,.
\ee

In the context of the Inverse Scattering Method \cite{KN,D,S1}
the separation variables $(u,v)$ appear usually as  pairs of canonically
conjugate variables sitting on the spectral curve of the related $n\times n$
Lax matrix $L(u)$. The coordinates $u_j$ are obtained respectively 
as the poles of the 
associated Baker-Akhiezer (BA) function $f(u)$ satisfying the linear problem
$$
L(u)\;f(u)=v\;f(u)\,,\qquad f(u)=(f_1(u),\ldots,f_n(u))^t\,,
$$
with some fixed normalisation $\vec\a(u)$
$$
\vec\a\cdot f\equiv \sum_{j=1}^n\a_j(u)\;f_j(u)=1\,.
$$
The method of SoV in such a formulation was successfully applied to many 
particular integrable systems, here are some of the relevant references
\cite{ES1,ES2,ES3,sl3,AHH,Scott,S1,KS1,Gekh,KS2,KS3,Toda,Factor}. 

In the present paper we prove the SoV for the classical $n$-particle 
Ruijsenaars system with the $n\times n$ Lax matrix found in \cite{Ru} 
and with the Hamiltonian
\beq
H_1=\sum_{j=1}^n \;\e^{p_j}\;
  \prod_{k\neq j}\frac{\s(x_j-x_k-\l)}{\s(x_j-x_k)}\,,\qquad \{p_j,x_k\}=
\delta_{jk}\,,
\eeq
where $\s(x)$ is the Weierstrass $\s$-function, $\l\in\R$ is a parameter 
of the model and $(p_j,x_j)$ are canonical Darboux variables. 
It is shown that the method of SoV applies to this system if we use
the standard normalisation vector $\vec\a$
\be
\vec\a=\vec\a_0\equiv (0,0,\ldots,0,1)\,, \quad \mbox{i.e.}\quad f_n(u)=1\,.
%\label{eq:a0}
\ee

The structure of the paper is the following. 
In Section 2 we collect known information 
about the Ruijsenaars system (Lax matrix, integrals of motion, etc). In Section
3 we give an overview of the method of separation of variables and apply
it then, in Section 4, to the system in question. In that key Section 
we also discuss the possibility of an alternative choice for the normalisation
vector $\vec\a(u)$. The generating functions of the canonical separating
transform
given in terms of the initial and separation variables are constructed 
in Section 5 
in explicit form for the case of two and three degrees of freedom. 
We also provide the separation of variables for the 
nonrelativistic limit $\l\rightarrow 0$ to the elliptic
Calogero-Moser system in Section 6.
The Section 7 contains some concluding remarks. 
%
%%%%%%%%%%%%%%%%%%%%%%%%%%%%%%  SECTION 2  %%%%%%%%%%%%%%%%%%%%%%%%%%%%%%%%%%%%%%%%
%
\newsection{The system}
\setcounter{equation}{0}
Let us first recall some properties of the Weierstrass functions
which we will need in the main text.
Let $2\omega_{1,2}\in\C$  be a fixed pair of the primitive periods and
$\Gamma=2\omega_1\Z+2\omega_2\Z$
the corresponding period lattice. Let us fix also the {\it primitive domain}
${\cal D}:=\{z=2\omega_1x+2\omega_2y\mid x,y\in[0,1)\}$ 
such that ${\cal D}\sim\C/\Gamma$.
The Weierstrass sigma-function
is defined by the infinite product (cf., for instance, \cite{WW})
\beq
\sigma(x) = x \prod_{\gamma\in\Gamma\setminus\{0\}} 
\left(1-\tfrac{x}{\gamma}
\right)\,
 \exp\left[ \tfrac{x}{\gamma} + \tfrac{1}{2}
( \tfrac{x}{\gamma})^2\right]\,,
\eeq
the relations between $\sigma$-, $\zeta$- and $\wp$- functions being given by
\beq
\zeta(x) =  \frac{\sigma^\prime(x)}{\sigma(x)}\,, \qquad
\wp(x) = - \zeta^\prime(x)\,,
\eeq
where $\sigma(x)$ and $\zeta(x)$ are odd functions and $\wp(x)$ is an 
even function of its argument.
We recall also that the $\sigma(x)$ is an entire function, and 
$\zeta(x)$ is a meromorphic function having simple poles at 
$\omega_{kl}$, both being quasi-periodic, obeying  
\[ 
\zeta(x+2\omega_{1,2}) = \zeta(x) + 2\,\eta_{1,2}\,,\qquad
\sigma(x+2\omega_{1,2}) = -\sigma(x)\;
\e^{2\eta_{1,2}(x+\omega_{1,2})}\,,
\] 
in which $\eta_{1,2}$ satisfy $\eta_1\,\omega_2 - \eta_2\,\omega_1 
= \frac{\pi i}{2}\,$, whereas $\wp(x)$ is doubly periodic.
{}From an algebraic point of view, the most important property of 
these functions is the existence of a number of functional 
relations, the most fundamental being 
\beq 
\zeta(\a) + \zeta(\b) + \zeta(\g) - \zeta(\a +\b + \g)
  = \frac{  \sigma(\a + \b )\,  \sigma(\b + \g )\, 
\sigma( \g + \a) }{ \sigma(\a)\, \sigma(\b) \,\sigma(\g)\, 
\sigma(\a + \b + \g )} 
\label{eq:zs}\eeq
which can be cast into the following form
\beq
\Phi_\kappa(x)\;\Phi_\kappa(y) = 
\Phi_{\kappa}(x+y)\;
\left[ \,\zeta(\kappa) +\zeta(x) +\zeta(y) -\zeta(\kappa+x+y)
\,\right] 
\label{eq:12} \eeq
with the function $\Phi_\kappa(x)$ defined as follows:
$$
\Phi_\kappa(x):=\frac{\s(x+\kappa)}{\s(x)\,\s(\kappa)}\,.
$$
Two other useful identities have the form
\bea
&&\Phi_{\kappa-\tilde\kappa}(a-b)\;\Phi_\kappa(x+b)\;\Phi_{\tilde\kappa}(y+a) 
- \Phi_{\kappa-\tilde\kappa}(x-y)\;\Phi_\kappa(y+a)\;
\Phi_{\tilde\kappa}(x+b) \qquad\qquad\label{zzz1} \\ 
&& ~ = \Phi_\kappa(x+a)\;\Phi_{\tilde\kappa}(y+b)\;\left[\; 
\zeta(a-b) + \zeta(x+b) - \zeta(x-y) - \zeta(y+a) \;\right]\,,\nonumber\eea  
\bea  
&&\Phi_{\kappa-\tilde\kappa}(x-y)\;\Phi_\kappa(y+a)\;\Phi_{\tilde\kappa}(x+a)
\qquad\qquad\label{zzz2}\\ 
&& ~ = \Phi_\kappa(x+a)\;\Phi_{\tilde\kappa}(y+a) \;\left[\; 
\zeta(x-y) - \zeta(\kappa+x+a) + \zeta(\tilde\kappa+y+a) + 
\zeta(\kappa-\tilde\kappa) \;\right]\,.  \nonumber
\eea 

The generalised Cauchy identity has the following form \cite{Fro}
\beq
\det\left( \Phi_\kappa(x_i - y_j)\right) = 
\Phi_\kappa( \Sigma )\; \sigma(\Sigma )\; 
\frac{ \prod_{k<l} \sigma(x_k - x_l) 
\,\sigma(y_l - y_k) }{  
\prod_{k,l} \sigma(x_k - y_l) }
\label{eq:cauchy} \eeq 
where $\Sigma\equiv \sum_i (x_i-y_i)$. 

Now we can introduce the $n$-particle ($A_{n-1}$ type) 
Ruijsenaars system \cite{Ru}. It is an integrable system 
with the following integrals of motion ($i=1,\ldots,n$)
\beq
H_i=\sum_{J\subset\{1,\ldots,n\} \atop \left|J\right|=i}
\exp\left(\sum_{j\in J}p_j\right)\,
  \prod_{j\in J \atop k\in \{1,\ldots,n\}\setminus J} 
\frac{\s(x_j-x_k-\l)}{\s(x_j-x_k)}\,.
\label{eq:def-H}\eeq
The variables $(p_j,x_j)$, $j=1,\ldots,n$, 
on a $2n$-dimensional symplectic manifold form
a canonical system, i.e. they possess the Poisson brackets
\beq
 \{p_j,x_k\}=\{x_j,x_k\}=0\,, \qquad \{p_j,x_k\}=\d_{jk}\,, 
 \qquad j,k=1,\ldots,n\,,
\label{eq:def-Weyl-can}
\eeq
or, equivalently, the symplectic form $\omega$ is expressed as
$\omega=\sum_j dp_j\wedge dx_j= d(\sum_jp_j$ $dx_j)$. The $\l$ is
a parameter of the model. This system was proposed by Ruijsenaars
as a relativistic analog of the Calogero-Moser system. 
\vskip 0.2cm

\noindent
{\bf Proposition 1 (\cite{Ru}).} {\it 
The Hamiltonians $H_j$ Poisson commute
\beq
 \{H_j,H_k\}=0, \qquad j,k=1,\ldots,n\,.
\label{eq:comm-H-cl}\eeq}
\vskip 0.2cm

The Lax matrix for this model has the form 
\beq
   L(u)=\sum_{i,j=1}^nh_i\;\Phi_u(x_i-x_j+\l)\;E_{ij}\,,
\qquad h_i:=\e^{p_i}\;\prod_{j\neq i}\frac{\s(x_i-x_j-\l)}{\s(x_i-x_j)}\,,
\label{eq:L}
\eeq
where the matrix $E_{ij}$ have the following entries: $(E_{ij})_{kl}=
\d_{ik}\,\d_{jl}$. 
Notice that Ruijsenaars \cite{Ru} used another gauge of the momenta such that two
are connected by the following canonical transformation:
\beq
p_i\rightarrow p_i+\log\prod_{j\neq i}
\sqrt{\frac{\s(x_i-x_j+\l)}{\s(x_i-x_j-\l)}}\,,
\qquad x_i\rightarrow x_i\,.
\label{canon}\eeq
\vskip 0.2cm

\noindent
{\bf Proposition 2 (\cite{Ru}).} {\it 
The characteristic polynomial of the matrix L(u) (\ref{eq:L})
generates the Hamiltonians (\ref{eq:def-H})
\beq
\det(L(u)-v\cdot\id)=\sum_{j=0}^n(-v)^{n-j}\;\frac{H_j}{\s^j(\l)}\;
\frac{\s(u+j\l)}{\s(u)}
\label{eq:char-poly-N}
\eeq
where we assume $H_0\equiv1$.}
%
%%%%%%%%%%%%%%%%%%%%%%%%%%%%%%  SECTION 3  %%%%%%%%%%%%%%%%%%%%%%%%%%%%%%%%%%%%%
%
\newsection{The method}
\setcounter{equation}{0}
Recall, first, the standard definitions of Liouville integrability and
SoV in the Hamil\-ton-Jacobi equation \cite{Ar}. 
An integrable Hamiltonian system with $N$ degrees of freedom is determined
by a $2N$-dimensional symplectic manifold (phase space) and $N$ independent
functions (Hamiltonians) $H_j$ commuting with respect to the Poisson bracket
\beq
 \{H_j,H_k\}=0\,,   \qquad j,k=1,\ldots,N\,.
\eeq

To find a SoV means then to find
a canonical transformation $M:(x,p)\mapsto (u,v)$,
$\;M:H_i(x,p)\mapsto H_i(u,v)$ such that there exist $N$ relations
\beq
\Phi_j(u_j,v_j;H_1,\ldots,H_N)=0\,,
\qquad j=1,\ldots,N\,,
\label{eq:sep-var-cl}
\eeq
separating the variables $u_j$.
The most common way to describe a canonical transformation is the one in terms 
of its generating function $F(u|x)$.

Presently, no algorithm is known of constructing a SoV for
any given integrable system. Nevertheless, 
there exists a fairly effective practical recipe
based on the classical inverse scattering method.
A detailed description of the procedure with many examples can be found in 
the review paper \cite{S1}, see also the works \cite{sl3,KS1,KS2,Toda,KS3}.
 Here we describe very briefly its main steps.

A Lax matrix for a given integrable system
is a matrix $L(u)$ dependent
on a ``spectral parameter'' $u\in \C$ such that its characteristic 
polynomial obeys two conditions
\bea
&&(i) \;\,\quad\mbox{{\it Poisson involutivity:}}\nonumber\\
&&\;\;\quad\quad \{\det(L(u)-v\cdot\id),\;\det(L(\tilde u)-\tilde v\cdot\id)\}=0\,,
\quad\forall u,\tilde u,v,\tilde v\in\C; \nonumber\\
&&(ii)\quad \det(L(u)-v\cdot\id) \quad \mbox{{\it generates all integrals
of motion}}\;\;H_i\,.
\nonumber\eea
A Baker-Akhiezer (BA) function is the eigenvector 
\beq
L(u)\;f(u)=v(u)\;f(u)
\label{5.5}\eeq
of the Lax matrix $L(u)$, provided that
a normalisation of the eigenvectors $f(u)$ is fixed
\beq
\vec\a\cdot f\equiv 
\sum_{i=1}^n\a_i(u)\;f_i(u)=1\,,\qquad (\;f(u)\equiv
(f_1(u),\ldots,f_n(u))^t\;)\,.
\label{5.6}\eeq
The pair $(u,v)$ can be thought of as a point of the {\it spectral curve} 
\beq
 \det(L(u)-v\cdot\id)=0\,.
\label{eq:spectral-curve}
\eeq
The BA function $f(u)$ is then a meromorphic function on the spectral curve.

The recipe for finding an SoV is simple:
\vskip 0.2cm
\noindent
{\it The separation variables $u_j$ are poles of the Baker-Akhiezer 
function, provided it is properly normalised.
The corresponding eigenvalues $v_j$ of $L(u_j)$, or some functions  
of them, serve as the canonically conjugated variables.
}
\vskip 0.2cm

It is easy to see that the pairs $(u_j,v_j)$ thus defined satisfy the 
separation equations \Ref{eq:sep-var-cl} for 
$\Phi_j\equiv\det(L(u_j)-v_j\cdot\id)$. The canonicity of the variables 
$(u_j,v_j)$ should be verified independently.
No general recipe is known how to guess the proper (that is producing
canonical variables) normalisation for the BA function. 
In many cases the simplest {\it standard} normalisation,
\beq
\vec\a(u)=\vec\a_0\equiv (0,0,\ldots,0,1)\,,
\label{eq:a0}\eeq
works. In other cases the vector $\vec\a$ may depend on the
spectral parameter $u$ and the dynamical variables $(x,p)$. We shall 
refer to such normalisation as a {\it dynamical} one.

{}From the linear problem \Ref{5.5} and normalisation \Ref{5.6} we derive that
$\vec\a\cdot L^k\;f=v^k\,,\;k=0,\ldots,n-1$, hence,
\beq
f=f(u)= \left( \begin{array}{c} \vec{\alpha}\cr \vec{\alpha}\cdot L(u) \cr
\vdots \cr \vec{\alpha}\cdot L^{n-1}(u) \end{array}\right)^{-1} \cdot 
\left(\begin{array}{c} 1\cr v\cr  \vdots \cr v^{n-1}\end{array}\right)\,.
\label{NN1}\eeq
Another useful representation of the eigenvector $f(u)$, which can be directly
verified, is as follows:
\beq
f_j(u)=\frac{(L(u)-v\cdot\id)^\wedge_{jk}}{(\vec\a\cdot(L(u)-v\cdot\id)^\wedge)_k}\,,
\qquad \forall k=1,\ldots,n\,,
\label{NN2}\eeq
where the wedge denotes the classical adjoint matrix
(matrix of cofactors).

To derive equations for the separation variables, let
$f_i^{(j)}=\res_{u=u_j}f_i(u)$ and $v_j\equiv v(u_j)$.
Then from (\ref{5.5})--(\ref{5.6}) we have the overdetermined system
of $n+1$ linear homogeneous equations for $n$ components $f_i^{(j)}$
of the vector $f^{(j)}$:
\beq
\left\{\begin{array}{l}
L(u_j)\;f^{(j)}=v_j\;f^{(j)}\,, \\
\sum_{i=1}^n\a_i(u_j)\;f_i^{(j)}=0\,.\end{array}\right.
\label{5.7}\eeq
The pair $(u,v)\equiv(u_j,v_j)$ is thus determined from the condition
\beq
\mbox{{\rm rank}}\pmatrix{\vec\a(u)\cr L(u)-v\cdot\id}=n-1\,.
\label{5.9}\eeq
Finally, the condition (\ref{5.9}) can be rewritten as the 
following vector equation:
\beq
\vec \a\cdot (L(u)-v\cdot\id)^\wedge=0\,.
\label{5.11}\eeq
One can eliminate $v$ from (\ref{5.11}) to get the equation for $u_j$'s
in the following way. From the linear system \Ref{5.7} it follows that 
$\vec\a\cdot(L(u_j))^k\,f^{(j)}=0$, $k=0,\ldots,n-1$, so that
(because $f^{(j)}$ is not a zero vector) the following determinant
has to vanish on the separation variables $u_j$:
\beq
B(u)=\det\pmatrix{\vec\a\cr
\vec\a \cdot L(u)\cr
\vdots\cr
\vec\a \cdot L^{n-1}(u)}=0\,.
\label{5.12}\eeq
The formula \Ref{5.12} for the separation variables
appeared already in \cite{Scott} (see also \cite{Gekh}) in the case of 
standard normalisation: $\vec\a=\vec\a_0$ \Ref{eq:a0} (see, for instance,
formula (22) in \cite{Scott}). 

Notice that the fact, that equations \Ref{5.11} and \Ref{5.12} are the ones 
for the poles of the BA function, is already hinted, respectively, by the formulas 
\Ref{NN2} and \Ref{NN1}.

Also, from equations (\ref{5.11}) we can get many various formulas for $v$ 
in the form
\beq
v=A(u)
\eeq
with $A(u)$ being rational functions of the entries of $L(u)$.
Let us describe those formulas for $A(u)$ explicitly. 

Define the matrices $L^{(p)}$, $p=1,\ldots,n$, with the following entries:
\beq
L^{(p)}_{ij}:=\sum_{i_1=1}^n\cdots \sum_{i_{p-1}=1}^n\;\left| 
\begin{array}{cccc}
L_{i,j}&L_{i,i_1}&\cdots & L_{i,i_{p-1}} \\ 
L_{i_1,j}&L_{i_1,i_1}&\cdots & L_{i_1,i_{p-1}} \\ 
\vdots &\vdots & \ddots & \vdots \\ 
L_{i_{p-1},j}& L_{i_{p-1},i_1} &\cdots & L_{i_{p-1},i_{p-1}}
\end{array}\right|\,,\quad p=2,3,\ldots,n\,,   
\label{eq:expA}\eeq
and put $L^{(1)}\equiv L$. These matrices satisfy the 
recursion relation of the form
\beq
L^{(p)}= L\left({\rm tr}\,L^{(p-1)}\right) - (p-1)\, L^{(p-1)} L\,.
\label{eq:recurs}\eeq 
Introduce the matrix ${\cal B}(u)$ by the formula
\beq
{\cal B}(u):=\pmatrix{
          \vec\a\cdot L^{(1)}(u)\;L^{-1}(u)\cr
          \vec\a\cdot L^{(2)}(u)\;L^{-1}(u)\cr
\tfrac12\;\vec\a\cdot L^{(3)}(u)\;L^{-1}(u)\cr
\cdots\cr
\tfrac{1}{(n-1)!}\;\vec\a\cdot L^{(n)}(u)\;L^{-1}(u)}\,.
\eeq
Then we have the following statement.
\vskip 0.2cm

\noindent
{\bf Proposition 3.} {\it
\beq
\vec\a\cdot(L(u)-v\cdot\id)^\wedge=((-v)^{n-1},(-v)^{n-2},\ldots,1)\cdot\,{\cal B}(u)\,.
\label{system}\eeq}
\vskip 0.2cm

\noindent
{\bf Proof.} 
The characteristic determinant $\det(L(u)-v\cdot\id)$ has the following
representation
\beq
\det(L(u)-v\cdot\id)=(-v)^n+\sum_{j=1}^n\frac{(-v)^{n-j}}{j!}\;\tr(L^{(j)}(u))\,.
\label{ewq}\eeq
The adjoint matrix $(L(u)-v\cdot\id)^\wedge$ is a matricial polynomial in $v$
of the degree $n-1$,
\beq
(L(u)-v\cdot\id)^\wedge=(-v)^{n-1}\cdot\id+\sum_{j=1}^{n-1}\,(-v)^{n-1-j}\;
A^{(j)}(u)\,.
\label{qwe}\eeq
In order to find the matrices $A^{(j)}$, substitute \Ref{ewq} and \Ref{qwe}
into the definition of the adjoint matrix,
\be
\det(L(u)-v\cdot\id)\cdot\id=
(L(u)-v\cdot\id)\;(L(u)-v\cdot\id)^\wedge\,,
\ee
and equate coefficients with the degrees of $v$. In this way we get 
the following recursion relation for the $A^{(j)}$'s 
\beq
A^{(j)}=\tfrac{1}{j!}\,\tr(L^{(j)})-L\,A^{(j-1)}
\eeq
with the initial data
\beq
A^{(0)}=\id\,,\qquad A^{(n-1)}=\tfrac{1}{n!}\,\tr(L^{(n)})\,L^{-1}(u)\,.
\eeq
The matrix $\tfrac{1}{j!}\,L^{(j+1)}\,L^{-1}$ (cf. \Ref{eq:recurs}) 
satisfies the same recursion and the same initial values which means that
$$
A^{(j)}(u)=\tfrac{1}{j!}\,L^{(j+1)}(u)\,L^{-1}(u)\,.
$$
\endproof
\vskip 0.2cm

{}From the system of linear homogeneous equations 
\beq
\vec\a\cdot(L(u)-v\cdot\id)^\wedge
\equiv((-v)^{n-1},(-v)^{n-2},\ldots,1)\cdot\,{\cal B}(u)=0
\eeq
(cf. \Ref{system}) we derive that
\beq
(-v)^{j-i}=\frac{({\cal B}^\wedge(u))_{ki}}{({\cal B}^\wedge(u))_{kj}}\,,
\qquad \forall k\,.
\label{super}\eeq
The formula \Ref{super} gives plenty of different representations for the 
function $A(u)$, all of them being compatible on the separation variables since,
because of the equality
\beq
B(u)=\det({\cal B}(u))\,,
\eeq
the matrix ${\cal B}^\wedge(u_j)$ has rank 1. 

To validate the choice of normalisation $\vec\a(u)$
it remains, first, to make  sure that the number of $u_j$'s is exactly $N$
(in some degenerate cases one has to supply a couple of extra variables to 
make a complete set) and, second, 
to verify (somehow) the canonicity of brackets between
the whole set of separation variables, namely: between zeros $u_j$ of $B(u)$ 
and their conjugated variables $v_j\equiv v(u_j)=A(u_j)$. 
To do this final
calculation one needs information about Poisson brackets between
entries of the Lax matrix $L(u)$. 
%
%%%%%%%%%%%%%%%%%%%%%%%%%%%%%%  SECTION 4  %%%%%%%%%%%%%%%%%%%%%%%%%%%%%%%%%%%%%%
%
\newsection{The separation}
\setcounter{equation}{0}
We now proceed with applying the general method to the system in question.
{}For the Ruijsenaars model the number $N$ of degrees of freedom 
coincides with the number $n$ of particles and, respectively, with 
the dimension $n$ of the Lax matrix (\ref{eq:L}), so we can put $N=n$ in the
formulas of the above Section. Let us first prove two useful Lemmas.
\vskip 0.2cm

\noindent
{\bf Lemma 1.} {\it Let $c_i\in\C$, $x^{(i)}_j\in{\cal D}$, $i=1,\ldots,M$,
$j=1,\ldots,N$, be arbitrary constants such that
\be
\sum_{j=1}^N x^{(i)}_j\equiv x\quad({\rm mod}\,\Gamma)\quad\qquad \forall i\,.
\ee
Then there exist $C\in\C$, $y_j\in{\cal D}$, $j=1,\ldots,N$, such that
\be
p(u)\equiv \sum_{i=1}^M c_i\prod_{j=1}^N\s(u-x^{(i)}_j)=
C\prod_{j=1}^N\s(u-y_j)\,,\qquad \forall u\in\C\,,
\ee
where
\be
\sum_{j=1}^N y_j\equiv x\quad({\rm mod}\,\Gamma)\,.
\ee}
\vskip 0.2cm

\noindent
The $p(u)$ can be thought of as $\s$-function version of the $N$th
degree polynomial (in $u$) which is represented in terms of its zeros
$y_j$.
\vskip 0.2cm

\noindent
{\bf Proof.}
Let $z_j\in{\cal D}$, $j=1,\ldots,N$, be $N$ distinct constants such that
$\sum_{j=1}^Nz_j\equiv x\quad({\rm mod}\,\Gamma)$. 
Consider the elliptic function $\tilde p(u)$ of the form
\beq
\tilde p(u)=\sum_{i=1}^M c_i\prod_{j=1}^N\frac{\s(u-x^{(i)}_j)}{\s(u-z_j)}\,.
\label{4}\eeq
Any elliptic function can be represented through the ratio of products
of $\s$-functions depending on its zeros, $y_j$, and its poles, $z_j$
(cf., for instance, \cite{Er}), i.e.
\beq
\tilde p(u)=C\prod_{j=1}^N\frac{\s(u-y_j)}{\s(u-z_j)}\,,
\label{5}\eeq
where $\sum_{j=1}^Ny_j\equiv\sum_{j=1}^Nz_j\equiv x\quad({\rm mod}\,\Gamma)$.
The statement follows if we equate
right hand sides of (\ref{4}) and (\ref{5}).
\endproof
\vskip 0.2cm

Consider the Lax matrix $L(u)$ for the Ruijsenaars system
\beq
L(u)=\sum_{i,j=1}^nh_i\;\Phi_u(x_i-x_j+\l)\;E_{ij}\,.
\label{66}\eeq
\newpage

%\vskip 0.2cm

\noindent
{\bf Lemma 2.} {\it For any integer $p=1,2,\ldots,n$ we have the identity
\beq
(L(u))^p+\sum_{j=1}^{p-1}(-1)^j\;\frac{H_j}{\s^j(\l)}\;
\frac{\s(u+j\l)}{\s(u)}\;
(L(u))^{p-j}
=\sum_{i,j=1}^nh_i\,C_{ij}^{(p)}\;\Phi_u(x_i-x_j+p\l)\;E_{ij}
\label{8}\eeq
where the scalars $C_{ij}^{(p)}$ do not depend on the spectral parameter $u$ 
and are given by the formula
\beq
C^{(p)}_{ij}=(-1)^{p-1}\sum_{i_1<\cdots <i_{p-1}}h_{i_1}\cdots h_{i_{p-1}}\;
\frac{ \prod_{k<l} \sigma(x_{i_k} - x_{i_l}) 
\,\sigma(x_{i_l} - x_{i_k}) }
{\prod_{k,l} \sigma(x_{i_k} - x_{i_l}+\l)}\;\times \qquad
\eeq
$$
\qquad\qquad\times\;
\frac{ \sigma(x_i-x_j + p\l)}{ \sigma(x_i-x_j + \l)}\;
\prod_{k=1}^{p-1}
\left[\frac{ \sigma(x_{i} - x_{i_k})\,\sigma(x_{i_k} - x_{j})}
{\sigma(x_{i} - x_{i_k}+\l)\,\sigma(x_{i_k} - x_{j}+\l)}\right]
$$
and $C^{(1)}_{ij}=1$.}
\vskip 0.2cm

\noindent
This Lemma does actually say that it is possible to arrange for the 
degree $p$ polynomial in $L(u)$ (the left hand side of (\ref{8}))
such that $u$-dependence of its $(ij)$-entry occurs only through
the factor $\Phi_u(x_i-x_j+p\l)$. This fact reflects some hidden
internal structure of the Lax matrix $L(u)$ and is essential
for further proof of the separation of variables. Notice also that
the usage of the generalised Cauchy identity is very important
for the proof of the Lemma given below. 
\vskip 0.2cm

\noindent
{\bf Proof.} Iterating the recursion \Ref{eq:recurs}
for the matrix $L^{(p)}(u)$, we get the formula
\beq
L^{(p)}=(-1)^{p-1}(p-1)!\,L^p+\sum_{j=1}^{p-1}(-1)^{p-1-j}\;\frac{(p-1)!}{j!}\;
\tr\,(L^{(j)})\;L^{p-j}\,.
\label{L-recurrence}\eeq
Noticing that the traces of the $L^{(j)}$ matrices are expressed in terms of 
the integrals of motion (cf. \Ref{eq:char-poly-N} and \Ref{ewq})
\beq
\tr\,L^{(j)}=j!\;\frac{H_j}{\s^j(\l)}\;\frac{\s(u+j\l)}{\s^j(u)}
\eeq
we have that
$$
h_i\,C_{ij}^{(p)}\;\Phi_u(x_i-x_j+p\l)=\frac{(-1)^{p-1}}{(p-1)!}
\;L^{(p)}_{ij}\,.
$$
The right hand side being evaluated with the help of the generalised
Cauchy identity (\ref{eq:cauchy}), we arrive to the statement of the Lemma.
\endproof
\vskip 0.2cm 

In order to separate variables in the Ruijsenaars system, first of all
we have to fix the normalisation vector $\vec\a(u)$. The crucial observation
is that we can use the standard normalisation \Ref{eq:a0}.
Then we have the following ``characteristic equations'' for the separation
variables $u=u_j$ and $v=v_j$ (cf. (\ref{5.11}))
\beq
{(L(u)-v\cdot\id)}^\wedge_{nk}=0\,, \qquad k=1,\ldots,n\,.
\label{55.11}\eeq
The ``$\s$-polynomial'' $B(u)$ (\ref{5.12}) has now the form
\beq
B(u)=\det\pmatrix{0&\ldots&1\cr
L_{n1}&\ldots&L_{nn}\cr
\vdots&\ddots&\vdots\cr
(L^{n-1})_{n1}&\ldots&(L^{n-1})_{nn}}\,.
\label{55.12}\eeq
Its zeros, $u_j$, are the poles of the BA function $f(u)$ and
are the separation variables. Let us first verify that we have got
the right number of the $u_j$'s. 
\vskip 0.2cm

\noindent
{\bf Theorem 1.} {\it $\s$-polynomial $B(u)$ (\ref{55.12})
has $n-1$ zeros $u_j\in{\cal D}$ and can be represented by the formula
\beq
B(u)=\tilde C\; \prod_{j=1}^{n-1}\Phi_u(-u_j)
\label{13}\eeq
where $\tilde C$ does not depend on the spectral parameter $u$
and has the form
\beq
\tilde C=(-1)^{n-1}\;h_n^{n-1}\;
\left|\matrix{1&\ldots&1\cr
C_{n1}^{(2)}&\ldots&C_{n,n-1}^{(2)}\cr
\vdots&\ddots&\vdots\cr
C_{n1}^{(n-1)}&\ldots&C_{n,n-1}^{(n-1)}}\right|\,.
\label{ccc}\eeq
Variables $u_j$ obey the restriction
\beq
\sum_{j=1}^{n-1}u_j\equiv\sum_{j=1}^{n-1}(x_j-x_n)-\tfrac{n(n-1)}{2}\;\l
\quad({\rm mod}\,\Gamma)\,.
\label{36}\eeq}
\vskip 0.2cm 

\noindent
{\bf Proof.} Using Lemma 2 we can represent $B(u)$ in the form
\beq
B(u)=(-1)^{n-1}\;h_n^{n-1}\times
\label{15}\eeq
\be
\times\left|\matrix{
\quad\Phi_u(x_{n1}+\l)
&\ldots&\quad\Phi_u(x_{n,n-1}+\l)\cr
C_{n1}^{(2)}\,\Phi_u(x_{n1}+2\l)&
\ldots&
C_{n,n-1}^{(2)}\,\Phi_u(x_{n,n-1}+2\l)\cr
\vdots&\ddots&\vdots\cr
C_{n1}^{(n-1)}\,\Phi_u(x_{n1}+(n-1)\l)&
\ldots&
C_{n,n-1}^{(n-1)}\,\Phi_u(x_{n,n-1}+(n-1)\l)}\right|\,.
\ee
Then, using Lemma 1, we conclude that the $\s$-polynomial $B(u)$
can be rewritten in terms of its zeros in the form \Ref{13} 
where $\tilde C$ is given by the formula \Ref{ccc} and we also 
have the restriction \Ref{36}.
\endproof
\vskip 0.2cm

To avoid discontinuities when discussing the Poisson brackets 
it is convenient to think of $u_j$'s as
lying on the torus $\C/\Gamma$ rather then on ${\cal D}$.

In the sequel we obtain few statements which are valid for a general
Lax matrix $L(u)$. Let us introduce the following matrices:
\beq
L(u,v):={L(u)-v\cdot\id}\,,\qquad
L^\wedge(u,v):={(L(u)-v\cdot\id)}^\wedge\,.
\eeq
We can express the Poisson brackets of $L^\wedge(u,v)$
with $L(\tilde u,\tilde v)$ in terms of the Poisson brackets
of $L(u,v)$ with $L(\tilde u,\tilde v)$. The answer is given 
by the following Lemma. 
\vskip 0.2cm

\noindent
{\bf Lemma 3.} {\it
\bea
\{L^\wedge_{1}(u,v),L_{2}(\tilde u,\tilde v)\}&=&
\Delta_1^{-1}\,\left(
\tr_1\,[\,L^\wedge_{1}(u,v)\,\{L_1(u,v),L_2(\tilde u,\tilde v)\}\,]
\right.\nonumber\\
&&\left.\qquad\qquad-\,L^\wedge_{1}(u,v)\,\{L_1(u,v),L_2(\tilde u,\tilde v)\}
\right)\,L^\wedge_{1}(u,v)\,,\nonumber\\
\{L_{1}(u,v),L^\wedge_{2}(\tilde u,\tilde v)\}&=&
\Delta_2^{-1}\,\left(
\tr_2\,[\,L^\wedge_{2}(\tilde u,\tilde v)\,
\{L_1(u,v),L_2(\tilde u,\tilde v)\}\,]
\right.\nonumber\\
&&\left.\qquad\qquad-\,L^\wedge_{2}(\tilde u,\tilde v)\,
\{L_1(u,v),L_2(\tilde u,\tilde v)\}\right)
\,L^\wedge_{2}(\tilde u,\tilde v)\,,
\nonumber\eea
where $L_{1}(u,v)=L(u,v)\otimes\id,\;L_{2}(\tilde u,\tilde v)=\id\otimes 
L(\tilde u,\tilde v),\;
L^\wedge_{1}(u,v)=L^\wedge(u,v)\otimes\id$ etc,
$\Delta_1=\det(L(u,v)),\;\Delta_2=\det(L(\tilde u,\tilde v))$
and $\tr_{1,2}$ means trace in the first, respectively, the second
space of the tensor product of two spaces and is defined by the rule:
\beq
\tr_1[A_1B_2]\equiv \tr_1[A\otimes B]:=\tr(A)\,(\id\otimes B)\equiv
\tr(A)\,B_2\,,
\eeq
\beq
\tr_2[A_1B_2]\equiv \tr_2[A\otimes B]:=\tr(B)\,(A\otimes\id)\equiv
\tr(B)\,A_1\,.
\eeq}
\vskip 0.2cm

\noindent
{\bf Proof.} The matrix $L(u,v)$ and its classical adjoint $L^\wedge(u,v)$
by definition satisfy the relation
\beq
L^\wedge(u,v)\,L(u,v)=L(u,v)\,L^\wedge(u,v)=\Delta_1\cdot\id\,.
\eeq
Differentiating this formula with respect to a parameter $t$ and using
the formula 
\be
\frac{d}{dt}\,\left(\det L\right)=\tr\,\left(L^\wedge\;\frac{d}{dt}\;L\right)
\ee
one obtains (cf. (1.45)--(1.47) from \cite{AHH})
\be
\frac{dL^\wedge}{dt}=\frac{
L^\wedge\,\tr\,\left(L^\wedge\;\frac{d}{dt}\;L\right)
-L^\wedge\,\left(\frac{d}{dt}\;L\right)\,L^\wedge}{\Delta_1}\,.
\ee
{}From which we have the following derivatives in the component-wise form:
\beq
\frac{\partial L^\wedge_{ij}}{\partial L_{pq}}=
\frac{L^\wedge_{qp}\,L^\wedge_{ij}-L^\wedge_{ip}\,L^\wedge_{qj}}{\Delta_1}\,.
\label{one-more}\eeq
Now, using the derivation property of the bracket,
\be
\{L^\wedge_{ij}(u,v),L_{kl}(\tilde u,\tilde v)\}=
\sum_{pq}\;\frac{\partial L^\wedge_{ij}(u,v)}{\partial L_{pq}(u,v)}\;
\{L_{pq}(u,v),L_{kl}(\tilde u,\tilde v)\}\,,
\ee
\be
\{L_{ij}(u,v),L^\wedge_{kl}(\tilde u,\tilde v)\}=
\sum_{pq}\;\frac{\partial L^\wedge_{kl}(\tilde u,\tilde v)}
{\partial L_{pq}(\tilde u,\tilde v)}\;
\{L_{ij}(u,v),L_{pq}(\tilde u,\tilde v)\}\,,
\ee
we verify both statements of the Lemma by substitution and straightforward 
calculation. 
\endproof
\vskip 0.2cm

{}From the involutivity of the characteristic polynomials of a Lax
matrix $L$, $\Delta_1$ and $\Delta_2$, we have the equality:
\bea
0&=&\{\Delta_1\cdot\id\otimes\id,\Delta_2\cdot\id\otimes\id\}=
\{L_{1}(u,v)\,L^\wedge_{1}(u,v),
L_{2}(\tilde u,\tilde v)\,L^\wedge_{2}(\tilde u,\tilde v)\}
\qquad\qquad
\nonumber\\
&=&L_{1}\,L_{2}\,\{L^\wedge_{1},L^\wedge_{2}\}+
L_{1}\,\{L^\wedge_{1},L_{2}\}\,L^\wedge_{2}
+L_{2}\,\{L_{1},L^\wedge_{2}\}\,L^\wedge_{1}+
\{L_{1},L_{2}\}\,L^\wedge_{1}\,L^\wedge_{2}\,.
\nonumber\eea
Hence, using Lemma 3, we can get from here an expression for the 
bracket of $L^\wedge$ with $L^\wedge$ in terms of the brackets
of $L$ with $L$.
\vskip 0.2cm

\noindent
{\bf Lemma 4.} 
\bea
\{L^\wedge_{1}(u,v),L^\wedge_{2}(\tilde u,\tilde v)\}
&=&\Delta_1^{-1}\,\Delta_2^{-1}\,
\left(\,L^\wedge_{1}\,L^\wedge_{2}\,
\{L_{1},L_{2}\}-
\tr_1\,[L^\wedge_{1}\,L^\wedge_{2}\,\{L_{1},L_{2}\}]\right.
\nonumber\\
&&\left.\qquad\qquad\qquad-\tr_2\,[L^\wedge_{1}\,L^\wedge_{2}\,
\{L_{1},L_{2}\}]\,\right)\;L^\wedge_{1}\,L^\wedge_{2}\,.\nonumber\eea
\vskip 0.2cm

Suppose now that a Lax matrix $L(u)$ satisfies the 
quadratic (dynamical) $(r,s)$-bracket, then we have the following statement.
\vskip 0.2cm

\noindent
{\bf Lemma 5.} {\it Let a Lax matrix $L(u)$ satisfy the quadratic
$(r,s)$-bracket of the form
\beq
\{L_{1}(u),L_{2}(\tilde u)\}=L_{1}\,L_{2}\,r_+-r_-\,L_{1}\,L_{2}
+L_{1}\,s_+\,L_{2}-L_{2}\,s_-\,L_{1}
\label{kaka}\eeq
where 
\be
r_+-r_-+s_+-s_-=0\,,\qquad {\cal P}
\;r_\pm\;{\cal P}=-{r_\pm}_{|_{u\leftrightarrow\tilde u}}\,,
\qquad {\cal P}\;s_\pm\;{\cal P}={s_\mp}_{|_{u\leftrightarrow\tilde u}}\,.
\ee
Here ${\cal P}$ is the flip in tensor product of two spaces, 
i.e. ${\cal P}\,(A\otimes B)\,{\cal P}=B\otimes A$.
Then the matrix $L^\wedge(u,v)\equiv{(L(u)-v\cdot\id)}^\wedge$ obeys
the bracket of the form
\bea
\{L^\wedge_{1},L^\wedge_{2}\}&=&
(r_+-\tr_1r_+-\tr_2r_+)\,L^\wedge_{1}L^\wedge_{2}-
L^\wedge_{1}L^\wedge_{2}\,(r_--\tr_1r_--\tr_2r_-)
\quad\label{xxx}\\
&+&L^\wedge_{2}\,(s_+-\tr_1s_+-\tr_2s_+)\,L^\wedge_{1}-
L^\wedge_{1}\,(s_--\tr_1s_--\tr_2s_-)\,L^\wedge_{2}
\nonumber\\
&+&v\;\Delta_1^{-1}\,\left[
(L^\wedge_{1}(r_+-s_-)-\tr_1[L^\wedge_{1}(r_+-s_-)])
\,L^\wedge_{1}L^\wedge_{2}\right.
\nonumber\\
&&\left.\qquad\qquad
\qquad\quad\quad-L^\wedge_{1}L^\wedge_{2}\,((r_--s_+)L^\wedge_{1}-
\tr_1[(r_--s_+)L^\wedge_{1}])\right]
\nonumber\\
&+&\tilde v\;\Delta_2^{-1}\,\left[
(L^\wedge_{2}(r_++s_+)
-\tr_2[L^\wedge_{2}(r_++s_+)])\,L^\wedge_{1}L^\wedge_{2}\right.
\nonumber\\
&&\left.\qquad\qquad
\qquad\quad\quad-L^\wedge_{1}L^\wedge_{2}\,((r_-+s_-)L^\wedge_{2}
-\tr_2[(r_-+s_-)L^\wedge_{2}])\right].
\nonumber\eea}
\vskip 0.2cm

\noindent
{\bf Proposition 4 (\cite{NKSR,Su}).} {\it The Lax matrix \Ref{eq:L}
of the Ruijsenaars model satisfies quadratic $(r,s)$-algebra 
\Ref{kaka} where $(r,s)$-matrices can be chosen to be as follows:
\beq
r_+=a-b+c-d\,,\qquad r_-=a+d\,,\qquad s_+=b+d\,,\qquad s_-=c-d\,,
\eeq
where
\bea
a&:=&\sum_{i\neq j}\,\Phi_{u-\tilde u}(x_i-x_j)\,E_{ij}\otimes E_{ji}
+\zeta(u-\tilde u)\,\sum_k\,E_{kk}\otimes E_{kk}\,,\label{N1}\\
b&:=&\sum_{i\neq j}\,\Phi_{u}(x_i-x_j)\,E_{ij}\otimes E_{ii}
+\zeta(u)\,\sum_k\,E_{kk}\otimes E_{kk}\,,\label{N2}\\
c&:=&\sum_{i\neq j}\,\Phi_{\tilde u}(x_i-x_j)\,E_{ii}\otimes E_{ij}
+\zeta(\tilde u)\,\sum_k\,E_{kk}\otimes E_{kk}\,,\label{N3}\\
d&:=&\sum_{i\neq j}\,\zeta(x_i-x_j)\,E_{ii}\otimes E_{jj}\,.\label{N4}\eea}
\vskip 0.2cm

\noindent
Notice here that one needs to use three algebraic relations 
\Ref{zzz1}--\Ref{zzz2} and \Ref{eq:12} for the function $\Phi$ 
to verify this $(r,s)$-structure (cf. \cite{NKSR}). 

Separation variables $(u,v)=(u_j,v_j)$, $j=1,\ldots,n-1$, for the 
Ruijsenaars model are implicitly defined by the following system
of equations 
\beq
(L^\wedge(u,v))_{nk}\equiv{(L(u)-v\cdot\id)}^\wedge_{nk}=0\,,
\qquad k=1,\ldots,n\,,
\label{the-system}\eeq
where $L(u)$ is the Lax matrix \Ref{eq:L}. The Poisson brackets for these
new variables are generally given by the expression:
\beq
\pmatrix{\{u_i,u_j\}&\{u_i,v_j\}\cr \{v_i,u_j\}&\{v_i,v_j\}}=
({\cal M}_{i;kl})^{-1}\times\qquad\qquad\qquad\qquad
\label{ppp}\eeq
\be
\times{\pmatrix{\{(L^\wedge(u,v))_{nk},(L^\wedge(\tilde u,\tilde v))_{nk}\}&
\{(L^\wedge(u,v))_{nk},(L^\wedge(\tilde u,\tilde v))_{nl}\}\cr
\{(L^\wedge(u,v))_{nl},(L^\wedge(\tilde u,\tilde v))_{nk}\}&
\{(L^\wedge(u,v))_{nl},(L^\wedge(\tilde u,\tilde v))_{nl}\}}}_{|_{A_{ij}}}\;
({\cal M}_{j;kl}^t)^{-1}
\ee
where it is assumed that $k\neq l$, the condition $A_{ij}$ means substitution
of the form
\be
A_{ij}:=\left\{\matrix{(u,v)=(u_i,v_i)\cr (\tilde u,\tilde v)=(u_j,v_j)}\right\}
\ee
and matrices ${\cal M}$ are defined as follows:
\beq
{\cal M}_{m;kl}:={\pmatrix{
\frac{\partial (L^\wedge(u,v))_{nk}}{\partial u}&
\frac{\partial (L^\wedge(u,v))_{nk}}{\partial v}\cr
\frac{\partial (L^\wedge(u,v))_{nl}}{\partial u}&
\frac{\partial (L^\wedge(u,v))_{nl}}{\partial v}}}_{|_{(u,v)=(u_m,v_m)}}\,.
\label{LL}\eeq
\vskip 0.2cm

\noindent
{\bf Theorem 2.} {\it The separation variables $(u_j,v_j)$, $j=1,\ldots,n-1$,
for the Ruijsenaars system,
defined by the system of equations \Ref{the-system}, possess the following 
Poisson brackets:
\bea
&&(i)\qquad \{u_i,u_j\}=\{u_i,v_j\}=\{v_i,v_j\}=0\,,\qquad i\neq j\,,\nonumber\\
&&(ii)\;\;\quad \{v_j,u_j\}=v_j\,.\nonumber\eea}
\vskip 0.2cm

\noindent
{\bf Proof.} Generically the matrix ${\cal M}_{m;kl}$ \Ref{LL} for $k\neq l$
is invertible
which means that in order to prove the statement $(i)$ we have to show that
\be
{\{(L^\wedge(u,v))_{nk},(L^\wedge(\tilde u,\tilde v))_{nl}\}}_{|_{A_{ij}}}=0\,,
\qquad \forall k,l=1,\ldots,n\,,
\ee
when $i\neq j$. The latter fact follows from the Lemma 5 when we substitute 
in the right hand side of \Ref{xxx} the $(r,s)$-matrices 
from the Proposition 4 and put in both sides $(u,v)=(u_i,v_i)$, $(\tilde u,
\tilde v)=(u_j,v_j)$, $i\neq j$. Indeed, using the definition \Ref{the-system},
we get then the expression of the form
\bea
&&{\{(L^\wedge(u,v))_{nk},(L^\wedge(\tilde u,\tilde v))_{nl}\}}_{|_{A_{ij}}}
\qquad\qquad\qquad\qquad\qquad\qquad(i\neq j)\label{XX}\\
&=&\sum_{pq}\;((a-b+c-d\,)_{np,nq})_{|_{A_{ij}}}\;(L^\wedge(u_i,v_i))_{pk}\;
(L^\wedge(u_j,v_j))_{ql}\nonumber\\
&+&\sum_{pqr} \left(\tfrac{v\;[(L^\wedge(u,v))_{np}(L^\wedge(u,v))_{qk}-
(L^\wedge(u,v))_{nk}(L^\wedge(u,v))_{qp}]}{\det(L(u,v))}\;
(a-b)_{pq,nr}\right)_{|_{A_{ij}}}\;
(L^\wedge(u_j,v_j))_{rl}\nonumber\\
&+&\sum_{pqr} \left(\tfrac{\tilde v\;
[(L^\wedge(\tilde u,\tilde v))_{np}(L^\wedge(\tilde u,\tilde v))_{ql}-
(L^\wedge(\tilde u,\tilde v))_{nl}(L^\wedge(\tilde u,\tilde v))_{qp}]}
{\det(L(\tilde u,\tilde v))}\;
(a+c)_{nr,pq}\right)_{|_{A_{ij}}}\;
(L^\wedge(u_i,v_i))_{rk}\,.\nonumber
\eea
Each of the above three terms is equal zero, the first one when simply inspecting
the inputs from the matrices $a,b,c,d$; the latter two because the simple zero
in the denominator is cancelled by a double zero in the numerator. 

In order to prove the statement $(ii)$ we take $i=j$ in \Ref{ppp}
to get
\beq
\{u_j,v_j\}\;\det\,{\cal M}_{j;kl}=
\{(L^\wedge(u,v))_{nk},(L^\wedge(u,v))_{nl}\}_{|_{(u,v)=(u_j,v_j)}}\,,
\label{the-last-one}\eeq
where we recall that $k\neq l$. Hence, we have to show that
\beq
-v_j\;\det\,{\cal M}_{j;kl}=
\{(L^\wedge(u,v))_{nk},(L^\wedge(u,v))_{nl}\}_{|_{(u,v)=(u_j,v_j)}}\,.
\label{tutu}\eeq
To calculate the right hand side of \Ref{tutu} we use the Proposition 4
and take the limit $\tilde u\rightarrow u$ in the $(r,s)$-bracket
\Ref{kaka}. Using the derivation property of the bracket and substituting 
$u=u_j,\,v=v_j$ we then conclude that the only non-vanishing term in the right
of \Ref{tutu} has the following form:
\be
\{(L^\wedge(u,v))_{nk},(L^\wedge(u,v))_{nl}\}_{|_{(u,v)=(u_j,v_j)}}
\qquad\qquad\qquad\qquad\qquad\qquad\qquad\qquad
\ee
\be
=v_j\;\sum_{prs}\left(\left(
\tfrac{\partial(L^\wedge(u,v))_{nk}}{\partial(L(u,v))_{pr}}\,
\tfrac{\partial(L^\wedge(u,v))_{nl}}{\partial(L(u,v))_{rs}}-
\tfrac{\partial(L^\wedge(u,v))_{nl}}{\partial(L(u,v))_{pr}}\,
\tfrac{\partial(L^\wedge(u,v))_{nk}}{\partial(L(u,v))_{rs}}\right)
\tfrac{\partial(L(u,v))_{ps}}{\partial u}\right)_{|_{(u,v)=(u_j,v_j)}}\,.
\ee
On the other hand the determinant of ${\cal M}_{j;kl}$ can be evaluated
making use of its definition \Ref{LL} and expressing the derivatives
by $u$ and $v$ in terms of those by $(L(u,v))_{pq}$. Then we have the following
formula for the left hand side of \Ref{tutu}:
\beq
-v_j\;\det\,{\cal M}_{j;kl}
\qquad\qquad\qquad\qquad\qquad\qquad\qquad\qquad\qquad\qquad
\label{XXX}\eeq
\be
=v_j\;\sum_{prs}\left(\left(
\tfrac{\partial(L^\wedge(u,v))_{nk}}{\partial(L(u,v))_{ps}}\,
\tfrac{\partial(L^\wedge(u,v))_{nl}}{\partial(L(u,v))_{rr}}-
\tfrac{\partial(L^\wedge(u,v))_{nl}}{\partial(L(u,v))_{ps}}\,
\tfrac{\partial(L^\wedge(u,v))_{nk}}{\partial(L(u,v))_{rr}}\right)
\tfrac{\partial(L(u,v))_{ps}}{\partial u}\right)_{|_{(u,v)=(u_j,v_j)}}\,.
\ee
Straightforward calculation, using \Ref{one-more} and the fact that
the matrix $L^\wedge(u_j,v_j)$ has rank 1
shows that these two expressions are equal to each other
(cf. here the proof of the analogous Theorem 1.3 from \cite{AHH} 
establishing the Poisson brackets for the separation variables  
for the $sl(n)$ Gaudin magnet obeying the simplest linear $r$-matrix algebra
with the rational $r$-matrix).
\endproof
\vskip 0.2cm

\noindent
{\bf Theorem 3.} {\it The variables $(u_j,y_j:=\log(v_j))$, $j=1,\ldots,n-1$,
together with the variables $(X,P)$ describing the ``motion of the 
center-of-mass'',
\beq
X:=x_n\,,\qquad P:=\log(H_n)=\sum_{j=1}^np_j\,,
\label{unvn}\eeq
constitute the complete canonical set of new (separation) variables.}
\vskip 0.2cm

\noindent
{\bf Proof.} The bracket $\{P,X\}=1$ is easily seen, so,
in addition to the statements of the Theorem 2, it is only left to check
that
\bea
&&\{P,u_j\}=\{P,v_j\}=0\,,\qquad j=1,\ldots,n-1\,,\label{a}\\
&&\{X,u_j\}=\{X,v_j\}=0\,,\qquad j=1,\ldots,n-1\,.\label{b}\eea
The equalities \Ref{a} are trivial since $(u_j,v_j)$, $j=1,\ldots,n-1$, 
are defined by the equations $(L(u)-v\cdot\id)^\wedge_{nk}=0$, $k=1,\ldots,n$, and
entries of the matrix $L(u)$ depend only on differences $x_i-x_j$, therefore
$$
\{P,\,(L(u))_{ij}\}=0\,,\qquad \forall i,j=1,\ldots,n\,.
$$
{}For the brackets in \Ref{b} we have the following expression ($k\neq l$):
\beq
{\cal M}_{j;kl}\;\pmatrix{\{X,u_j\}\cr\{X,v_j\}}=
-\pmatrix{\{X,(L^\wedge(u,v))_{nk}\}\cr
\{X,(L^\wedge(u,v))_{nl}\}}_{|_{(u,v)=(u_j,\,v_j)}}\,.
\label{4X}\eeq
The vector on the right of \Ref{4X} is equal to zero since $\forall 
k=1,\ldots,n$
$$
{\{X,(L^\wedge(u,v))_{nk}\}}_{|_{\tiny
\matrix{u=u_j\cr v=v_j}}}
=-\left[\sum_{pq}\;\tfrac{L^\wedge_{nk}L^\wedge_{qp}-
L^\wedge_{np}L^\wedge_{qk}}{\det(L(u,v))}\;
\delta_{np}\;(L_{pq}+v\,\delta_{pq})\right]_{|_{\tiny
\matrix{u=u_j\cr v=v_j}}}=0\,.
$$
The equalities \Ref{b} follow because the matrix ${\cal M}_{j;kl}$ 
is nondegenerate. 
\endproof
\vskip 0.2cm

The proved SoV for the $A_{n-1}$ ($n$-particle) problem with the standard
normalisation vector $\vec\a_0\equiv (0,0,\ldots,0,1)$ actually 
implies another SoV for the $A_{n-2}$ problem with the non-standard
normalisation vector $\vec\a_1$:
\beq
\vec\a_1:=(\,\Phi_u(\xi-x_1+\l),\;\ldots,\;\Phi_u(\xi-x_{n-1}+\l)\,)\,,
\label{eq:a1}\eeq
if we choose $\xi=x_n$. Let us demonstrate this explicitly. 

Let us take the Lax matrix \Ref{eq:L} for the $n$-particle system
\be
L(u)=\sum_{i,j=1}^nh_i\;\Phi_u(x_i-x_j+\l)\;E_{ij}\,.
\ee
If we remove the last ($n$th) row and the last column from this Lax matrix
then we get the following $(n-1)\times (n-1)$ matrix
\beq
L_{\times n}^{\times n}(u):=\pmatrix{
h_1\,\Phi_u(\l)&\cdots&h_1\,\Phi_u(x_1-x_{n-1}+\l)\cr
\vdots&\ddots&\vdots\cr
h_{n-1}\,\Phi_u(x_{n-1}-x_1+\l)&\cdots&h_{n-1}\,\Phi_u(\l)}
\eeq
which is the Lax matrix for the integrable system with $n-1$ particles
with the Hamiltonian
\beq
H_1^{(\times n)}=\Phi_u(\l)\;\sum_{i=1}^{n-1}h_i=\Phi_u(\l)\;\sum_{i=1}^{n-1}
\;\e^{p_i}\,\prod_{k\neq i}^n\,\frac{\s(x_i-x_k-\l)}{\s(x_i-x_k)}\,.
\label{4H}\eeq
Under the simple canonical transformation, 
\beq
\e^{p_i}\rightarrow \e^{p_i}\;\frac{\s(x_i-x_n)}{\s(x_i-x_n-\l)}\,,
\qquad x_i\rightarrow x_i\,,\qquad i=1,\ldots,n-1\,,
\eeq
the system \Ref{4H} turns into Ruijsenaars' system with $n-1$ particles. 
This 1-degree-of-freedom-less system obviously inherits 
the non-standard SoV with the dynamical normalisation (\ref{eq:a1})
from the standard one (with $\vec\a_0$) for the system with $n$ degrees
of freedom. Indeed, to see this, it is sufficient to note that the 
separation variables $(u_j,v_j)$, $j=1,\ldots,n-1$, 
for {\it both} systems are defined from the intersection
of two spectral curves:
\beq
\left\{\matrix{\det(L(u)-v\cdot\id)=0\,,\cr 
\det(L_{\times n}^{\times n}(u)-v\cdot\id)=0\,.}\right.
\label{5kk}\eeq
In other words, the condition of the standard SoV for the first problem,
\beq
\mbox{{\rm rank}}\pmatrix{\vec\a_0\cr L(u)-v\cdot\id}=n-1\,,
\label{6kk}\eeq
implies the following condition of SoV for the second problem:
\beq
\mbox{{\rm rank}}\pmatrix{\vec\a_1\cr L_{\times n}^{\times n}(u)-v\cdot\id}=n-2\,,
\label{7kk}\eeq
where $\vec\a_1(u)$ is given by \Ref{eq:a1}.

Procedure shown above, on how to connect the standard normalisation vector 
$\vec\a_0$ and the alternative one, $\vec\a_1$, does obviously reflect
an embedding, $gl(n-1)\subset gl(n)$, of one problem into the other.
In other words (and it is true in general, for any integrable
system of $A_n$ type), one always has a free choice, namely:
to include or not to include the ``center-of-mass variable'', $X$,
and its conjugate one, $P$, in the complete set of separation variables. 
%
%%%%%%%%%%%%%%%%%%%%%%%%%%%%%%  SECTION 5  %%%%%%%%%%%%%%%%%%%%%%%%%%%%%%%%%%%%%%%%
%
\newsection{Generating functions}
\setcounter{equation}{0}
In this Section we derive the explicit formulas for SoV in the simplest
cases: $n=3$ with the standard normalisation \Ref{eq:a0} of $\vec\a$, and
$n=2$ with the dynamical normalisation \Ref{eq:a1}
(we skip the trivial case of the purely coordinate SoV 
$x_{1,2}\rightarrow x_1\pm x_2$ for the $2$-particle problem).
Since the both cases are
treated in very much the same manner as their trigonometric
prototypes, see, respectively \cite{KS2} and \cite{KS3},
we present only the main formulas here, omitting
the details of the calculations.

Let us start with the $n=3$ case.
Following \cite{KS2} define two functions $A_1(u)$ and $A_2(u)$ by the
formulas
\beq
(L(u)-A_k)^\wedge_{3,3-k}=0\,,\qquad k=1,2\,,
\label{QQ}\eeq
or explicitly,
\beq
  A_k(u)=L_{kk}-\frac{L_{3k}\,L_{k,3-k}}{L_{3,3-k}}=e^{p_k}\,a_k(u)\,, \qquad
  k=1,2\,,
\eeq
\beq
  a_k(u)=\frac{\s(u+2\l+x_3-x_{3-k})\,\s(x_k-x_{3-k}-\l)}%
{\s(\l)\,\s(u+\l+x_3-x_{3-k})\,\s(x_k-x_{3-k}+\l)}\,, \qquad
k=1,2\,.\eeq
The separated variables $u_j$ are defined from the equation
\beq
  A_1(u_j)=A_2(u_j)
\eeq
which is equivalent to the equation $B(u_j)=0$ since
$$
B(u)=h_3^2\,\Phi_u(x_3-x_1+\l)\,\Phi_u(x_3-x_2+\l)\,\s(\l)\;(A_2-A_1)
$$
and has two roots $u_{1,2}\in{\cal D}$.
{}From the easily verified invariance of the ratio $a_1(u)/a_2(u)$ under the
transformation $u\mapsto x_1+x_2-2x_3-3\l-u$ it follows that
\beq
 u_1+u_2\equiv x_1+x_2-2x_3-3\l \quad({\rm mod}\,\Gamma)\,,
\label{eq:constraint}
\eeq
which agrees with (\ref{36}).
The conjugated variables $v_j\equiv e^{y_j}$ are defined as
\beq
  v_j=A_1(u_j)=A_2(u_j)
\eeq
or, equivalently, through four equations
\beq
 v_j=e^{p_k}\,a_k(u_j)\,, \qquad j,k\in\{1,2\}\,,
\eeq
for four variables $u_1$, $u_2$, $v_1$, $v_2$.
By virtue of the Theorem 3 the variables $(u_1,u_2,X;$ $y_1,y_2,P)$ 
are canonical.
The generating function of the separating canonical transformation $M$ is most
conveniently expressed in terms of another set of canonical variables
\beq
 x_+=x_1+x_2-2x_3\,,\qquad
 x_-=x_1-x_2\,, \qquad
 X=x_3\,,
\label{xFF}\eeq
\beq
 p_\pm=\tfrac12\,(p_1\pm p_2)\,, \qquad P=p_1+p_2+p_3\,,
\eeq
\beq
 u_\pm=u_1\pm u_2\,, \qquad y_\pm=\tfrac12\,(y_1\pm y_2)\,.
\label{xxFF}\eeq

We shall need a $\sigma$-generalisation of the Euler dilogarithm function, 
$$
{\rm Li}_2(z)=\int_0^z \log(\sin(\zeta))\;d\zeta\,,
$$
which we define as
\beq
S(z):=\int_0^z \log(\sigma(\zeta))\;d\zeta\,. 
\label{SS}\eeq
Notice that this function was introduced in \cite{nrk94} and has been used 
to construct 
the Lagrangian function of the integrable map which is a time-discretisation
of the Ruijsenaars system. 
Using the product expansion for the Weierstrass sigma-function
($q=\exp(i\pi\tfrac{\omega_2}{\omega_1})$):
\beq
\sigma(z)=\frac{2\omega_1}{\pi}\;\e^{\tfrac{\eta_1 z^2}{2\omega_1}}
\sin\left(\frac{\pi z}{2\omega_1}\right)\prod_{n=1}^\infty
\left( \frac{ 1-2q^{2n}\cos\left(\frac{\pi z}{\omega_1}\right)+q^{4n}
}{(1-q^{2n})^2}\right)\,,  
\eeq
cf. \cite{WW}, we can express the function $S$ in terms of
the following function
\beq
{\rm Li}_3(z;q):= \sum_{k=1}^\infty \frac{z^k}{(1-q^k)\;k^2}\,,
\qquad |q|<1\,,\quad |z|<1\,. 
\label{eq:qLi}\eeq
Notice that similar, but different from \Ref{eq:qLi}, $q$-deformations 
of the Euler (di-) trilogarithm have been proposed in the
review article \cite{Kir}. In terms of (\ref{eq:qLi}) we obtain
\beq
S(z)=\frac{\eta_1 z^3}{6\omega_1} 
+\left(\log\left(\frac{2\omega_1}{\pi}\right)
+2\sum_{k=1}^\infty\,\frac{q^{2k}}{(1-q^{2k})k}\right)\;z 
\eeq
\be
+ \frac{i\omega_1}{\pi}\left( 
{\rm Li}_3(q^2t;q^2) - {\rm Li}_3(q^2t^{-1};q^2)\right)\,,
\ee
where $t=\exp(\pi iz/\omega_1)$. This series representation converges
for $|q|^2\leq |t|\leq |q|^{-2}$. 

Let
\beq
 {\cal L}(\nu;x,y):=
 S(\nu+x+y)+S(\nu-x+y)+S(\nu+x-y)+S(\nu-x-y)\,.
\eeq
The generating function $F(y_+,x_+;u_-,x_-)$ of the canonical transformation
from $(x_\pm,p_\pm)$ to $(u_\pm,y_\pm)$, satisfying the defining relations
\beq
 \frac{\dd F}{\dd x_+}=p_+\,, \qquad
 \frac{\dd F}{\dd y_+}=u_+\,, \qquad
 \frac{\dd F}{\dd x_-}=p_-\,, \qquad
 \frac{\dd F}{\dd u_-}=-y_-\,,
\eeq
is given then by the expression
\begin{eqnarray}
 F&=&y_+\,(x_+-3\l)+x_+\,\log\s(\l)
     -{\cal L}\left(\frac\l2\,;\frac{x_-}{2}\,,\frac{u_-}{2}\right)
     \label{FF} \\
 && +\,S(\l-x_-)+S(\l+x_-)\,.
\nonumber\end{eqnarray}

The case $n=2$ with the normalisation 
$$
\vec\a_1=(\,\Phi_u(\xi-x_1+\l),\;\Phi_u(\xi-x_2+\l)\,)
$$
(cf. \Ref{eq:a1})
is treated similarly to its trigonometric prototype \cite{KS3}.
Having introduced the functions $A_1(u)$ and $A_2(u)$ by the formulas
\beq
(\vec\a_1\cdot(L(u)-A_k)^\wedge)_{3-k}=0\,,\qquad k=1,2\,,
\label{QQQ}\eeq
or explicitly,
\beq
 A_1=L_{11}-\tfrac{\Phi_u(\xi-x_1+\l)}{\Phi_u(\xi-x_2+\l)}\;L_{12}
=e^{p_1}\,a_1(u)\,,
\eeq 
\beq
 A_2=L_{22}-\tfrac{\Phi_u(\xi-x_2+\l)}{\Phi_u(\xi-x_1+\l)}\;L_{21}
=e^{p_2}\,a_2(u)\,,
\eeq
\beq
 a_k(u)=\tfrac{\s(u+\xi+2\l-x_{3-k})\,\s(\xi-x_k)\,\s(x_k-x_{3-k}-\l)}%
{\s(\l)\,\s(u+\xi+\l-x_{3-k})\,\s(\xi+\l-x_k)\,\s(x_k-x_{3-k}+\l)}\,, \quad
k=1,2\,,
\eeq
one proceeds as above with the only difference that the relation 
\Ref{eq:constraint} is replaced by
\beq
 u_1+u_2\equiv x_1+x_2-3\l-2\xi \quad ({\rm mod}\,\Gamma)
\eeq
and the variables $x_\pm$ are defined now as $x_\pm=x_1\pm x_2$.
The resulting expression for $F(y_+,x_+;u_-,x_-)$ is
\begin{eqnarray}
 F&=&y_+\,(x_+-3\l-2\xi)+x_+\,\log\s(\l) \label{FFF} \\
&& -\,{\cal L}\left(\frac\l2\,;\frac{x_-}{2}\,,\frac{u_-}{2}\right)
   -{\cal L}\left(\frac\l2\,;\frac{x_+-\l}{2}-\xi,\frac{x_-}{2}\right)
     \nonumber \\
 && +\,S(\l-x_-)+S(\l+x_-)\,.
\nonumber\end{eqnarray}
%
%%%%%%%%%%%%%%%%%%%%%%%%%%%%%%  SECTION 6  %%%%%%%%%%%%%%%%%%%%%%%%%%%%%%%%%%%%%%%%
%
\newsection{Nonrelativistic limit to the Calogero-Moser system}
\setcounter{equation}{0}
The nonrelativistic limit is obtained by letting $\l\rightarrow 0$ 
while rescaling the momenta $p_j:=i\l p_j/g$, $g\in\R$, and making the canonical
transformation $p_j:=p_j-ig\sum_{k\neq j}\zeta(x_j-x_k)$
such that $h_j\rightarrow 1+i\l\, p_j/g+O(\l^2)$
in (\ref{eq:L}). The $(r,s)$-matrix structure is linear in that limit 
since the $L$-matrix behaves as 
\beq
L(u)\;\rightarrow\  \ (\,\l^{-1}+\zeta(u)\,)\cdot\id + \tfrac{i}{g}\;
\ell(u) +  O(\l)\,,  
\label{LLL}\eeq
\beq
\ell(u):=\sum_j p_j\, E_{jj} -ig \sum_{j\neq k} \Phi_u(x_j-x_k)\, E_{jk}\,.
\label{LLLL}\eeq
The $\ell$-matrix \Ref{LLLL} is Krichever's \cite{Kr} 
Lax operator for the elliptic
Calogero-Moser system with the Hamiltonian
\beq
H=\sum_{j=1}^np_j^2+g^2\sum_{j\neq k}\wp(x_j-x_k)\,.
\eeq
\vskip 0.2cm

\noindent
{\bf Proposition 5 (\cite{ES4}).} {\it The Lax matrix $\ell(u)$ \Ref{LLLL}
of the elliptic Calogero-Moser system satisfies linear $(r,s)$-algebra 
of the form
\beq
\{\ell_1(u),\ell_2(\tilde u)\}=[\ell_1,r]+[\ell_2,s]
\label{skobka}\eeq
where 
\beq
r=a+c\,,\qquad s=a-b\,,\qquad 
s=-{\cal P}\;r\;{\cal P}_{|_{u\leftrightarrow \tilde u}}\,,
\eeq
(see \Ref{N1},\Ref{N2},\Ref{N3}),
and $[\,.\,,.\,]$ means matrix commutator.}
\vskip 0.2cm

The SoV for the elliptic Calogero-Moser system follows, in principle,
by taking limit $\l\rightarrow 0$ in the corresponding formulas 
describing SoV for the Ruijsenaars system. Although, because this limit 
is not so simple and straightforward, we prefer to do it independently,
repeating the steps for proving main statements for the Ruijsenaars
system in Section 4. 

The normalisation vector is the same:
$$
\vec\a(u)=\vec\a_0\equiv (0,0,\ldots,0,1)\,.
$$
We have now the following characteristic equations for the separation
variables $u=u_j$ and $v=v_j$ 
\beq
{(\ell(u)-v\cdot\id)}^\wedge_{nk}=0\,, \qquad k=1,\ldots,n\,.
\label{x55.11}\eeq
The zeros of the $\s$-polynomial $b(u)$ 
\beq
b(u):=\det\pmatrix{0&\ldots&1\cr
\ell_{n1}&\ldots&\ell_{nn}\cr
\vdots&\ddots&\vdots\cr
(\ell^{n-1})_{n1}&\ldots&(\ell^{n-1})_{nn}}
\label{x55.12}\eeq
give us separation variables $u_j$. 
\vskip 0.2cm

\noindent
{\bf Theorem 4.} {\it $\s$-polynomial $b(u)$ (\ref{x55.12})
has $n-1$ zeros $u_j\in{\cal D}$ and can be represented by the formula
\beq
b(u)=\tilde C\; \prod_{j=1}^{n-1}\Phi_u(-u_j)
\label{x13}\eeq
where $\tilde C$ does not depend on the spectral parameter $u$.
Variables $u_j$ obey the restriction
\beq
\sum_{j=1}^{n-1}u_j\equiv\sum_{j=1}^{n-1}(x_j-x_n)\quad ({\rm mod}\,\Gamma)\,.
\label{x36}\eeq}
\vskip 0.2cm 

\noindent
{\bf Proof.} From the limit \Ref{LLL} and the definitions of $B(u)$ 
and $b(u)$ we conclude that
$$
B(u)=b(u)+O(\l)\,.
$$
Both $B(u)$ and $b(u)$ are $\s$-polynomials in $u$ and, since the degree
of such a polynomial must not change with the analytical continuation of the 
parameter $\l$, $b(u)$ has the same degree as $B(u)$ does. Moreover,
now the separation variables have to obey the restriction \Ref{x36},
the one being the limit of the corresponding relation \Ref{36}.
\endproof
\vskip 0.2cm

Let us introduce the following notations:
\beq
\ell(u,v):={\ell(u)-v\cdot\id}\,,\qquad
\ell^\wedge(u,v):={(\ell(u)-v\cdot\id)}^\wedge\,,
\eeq
and also $\Delta_1=\det(\ell(u,v))$ and $\Delta_2=\det(\ell(\tilde u,\tilde v))$.
Suppose now that a Lax matrix $\ell(u)$ satisfies the 
linear (dynamical) $(r,s)$-bracket \Ref{skobka}, 
then we have the following statement.
\vskip 0.2cm

\noindent
{\bf Lemma 6.} {\it Let a Lax matrix $\ell(u)$ satisfy the linear 
$(r,s)$-bracket of the form
\beq
\{\ell_1(u),\ell_2(\tilde u)\}=[\ell_1,r]+[\ell_2,s]\,,
\qquad s=-{\cal P}\;r\;{\cal P}_{|_{u\leftrightarrow \tilde u}}\,.
\label{skobka1}\eeq
Then the matrix $\ell^\wedge(u,v)\equiv{(\ell(u)-v\cdot\id)}^\wedge$ obeys
the bracket of the form
\bea
\{\ell^\wedge_{1},\ell^\wedge_{2}\}
&=&\Delta_1^{-1}\;\left[\;(\,\ell_1^\wedge\,s-\tr_1\,[\,\ell_1^\wedge\,s\,]\,)\;
\ell_1^\wedge\,\ell_2^\wedge-\ell_1^\wedge\,\ell_2^\wedge\;
(\,s\,\ell_1^\wedge-\tr_1\,[\,s\,\ell_1^\wedge\,]\,)\;\right]
\qquad\label{skobka2}\\
&+&\Delta_2^{-1}\;\left[\;(\,\ell_2^\wedge\,r-\tr_2\,[\,\ell_2^\wedge\,r\,]\,)\;
\ell_1^\wedge\,\ell_2^\wedge-\ell_1^\wedge\,\ell_2^\wedge\;
(\,r\,\ell_2^\wedge-\tr_2\,[\,r\,\ell_2^\wedge\,]\,)\;\right]\,.\nonumber\eea}
\vskip 0.2cm

\noindent
{\bf Theorem 5.} {\it The separation variables $(u_j,v_j)$, $j=1,\ldots,n-1$,
for the elliptic Calogero-Moser system, defined by the system of equations 
\Ref{x55.11}, possess the following Poisson brackets:
\bea
&&(i)\qquad \{u_i,u_j\}=\{u_i,v_j\}=\{v_i,v_j\}=0\,,\qquad i\neq j\,,\nonumber\\
&&(ii)\;\;\quad \{v_j,u_j\}=1\,.\nonumber\eea}
\vskip 0.2cm

\noindent
{\bf Proof.} In analogy with the proof of the Theorem 2 we have to show
first that 
\be
{\{(\ell^\wedge(u,v))_{nk},(\ell^\wedge(\tilde u,\tilde v))_{nl}\}}_{|_{A_{ij}}}=0\,,
\qquad \forall k,l=1,\ldots,n\,,
\ee
when $i\neq j$. We have from Lemma 6 and Proposition 5 that
\bea
&&{\{(\ell^\wedge(u,v))_{nk},(\ell^\wedge(\tilde u,\tilde v))_{nl}\}}_{|_{A_{ij}}}
\qquad\qquad\qquad\qquad\qquad\qquad(i\neq j)\label{XX4}\\
&=&\left[
\Delta_1^{-1}\;(\,\ell_1^\wedge\,s-\tr_1\,[\,\ell_1^\wedge\,s\,]\,)\;
\ell_1^\wedge\,\ell_2^\wedge+\Delta_2^{-1}\;(\,\ell_2^\wedge\,r
-\tr_2\,[\,\ell_2^\wedge\,r\,]\,)\;
\ell_1^\wedge\,\ell_2^\wedge\right]_{|_{A_{ij}}}\nonumber\\
&=&\sum_{pqr}\left[\tfrac{(\ell^\wedge(u,v))_{np}(\ell^\wedge(u,v))_{qk}-
(\ell^\wedge(u,v))_{nk}(\ell^\wedge(u,v))_{qp}}{\det(\ell(u,v))}\;
(a-b)_{pq,nr}\right]_{|_{A_{ij}}}\;
(\ell^\wedge(u_j,v_j))_{rl}\nonumber\\
&+&\sum_{pqr} \left[\tfrac{
(\ell^\wedge(\tilde u,\tilde v))_{np}(\ell^\wedge(\tilde u,\tilde v))_{ql}-
(\ell^\wedge(\tilde u,\tilde v))_{nl}(\ell^\wedge(\tilde u,\tilde v))_{qp}}
{\det(\ell(\tilde u,\tilde v))}\;
(a+c)_{nr,pq}\right]_{|_{A_{ij}}}\;
(\ell^\wedge(u_i,v_i))_{rk}\,.\nonumber
\eea
These two terms in the right hand side have the same form as latter two
in \Ref{XX} and, again, they are equal to zero since in both
expressions the simple zero in the denominator is cancelled by
a double zero in the numerator. 

The matrix of derivatives ${\cal M}$ instead of \Ref{LL} has now the form
\beq
{\cal M}_{m;kl}:={\pmatrix{
\frac{\partial (\ell^\wedge(u,v))_{nk}}{\partial u}&
\frac{\partial (\ell^\wedge(u,v))_{nk}}{\partial v}\cr
\frac{\partial (\ell^\wedge(u,v))_{nl}}{\partial u}&
\frac{\partial (\ell^\wedge(u,v))_{nl}}{\partial v}}}_{|_{(u,v)=(u_m,v_m)}}\,.
\label{LL1}\eeq
In order to prove the statement $(ii)$ we have to show that
\beq
-\det\,{\cal M}_{j;kl}=
\{(\ell^\wedge(u,v))_{nk},(\ell^\wedge(u,v))_{nl}\}_{|_{(u,v)=(u_j,v_j)}}\,,
\label{the-last-one1}\eeq
where $k\neq l$. Again, the right hand side of \Ref{the-last-one1} can be evaluated
by first taking the limit $\tilde u \rightarrow u$ in the $(r,s)$-bracket
of the Proposition 5 and then using the derivation property of the 
bracket. We derive the following expression
\be
\{(\ell^\wedge(u,v))_{nk},(\ell^\wedge(u,v))_{nl}\}_{|_{(u,v)=(u_j,v_j)}}
\qquad\qquad\qquad\qquad\qquad\qquad\qquad\qquad
\ee
\be
=\sum_{prs}\left(\left(
\tfrac{\partial(\ell^\wedge(u,v))_{nk}}{\partial(\ell(u,v))_{pr}}\,
\tfrac{\partial(\ell^\wedge(u,v))_{nl}}{\partial(\ell(u,v))_{rs}}-
\tfrac{\partial(\ell^\wedge(u,v))_{nl}}{\partial(\ell(u,v))_{pr}}\,
\tfrac{\partial(\ell^\wedge(u,v))_{nk}}{\partial(\ell(u,v))_{rs}}\right)
\tfrac{\partial(\ell(u,v))_{ps}}{\partial u}\right)_{|_{(u,v)=(u_j,v_j)}}\,.
\ee
On the other hand the determinant of ${\cal M}_{j;kl}$ (cf. \Ref{XXX})
has the form
\beq
-\det\,{\cal M}_{j;kl}
\qquad\qquad\qquad\qquad\qquad\qquad\qquad\qquad\qquad\qquad
\label{XXX1}\eeq
\be
=\sum_{prs}\left(\left(
\tfrac{\partial(\ell^\wedge(u,v))_{nk}}{\partial(\ell(u,v))_{ps}}\,
\tfrac{\partial(\ell^\wedge(u,v))_{nl}}{\partial(\ell(u,v))_{rr}}-
\tfrac{\partial(\ell^\wedge(u,v))_{nl}}{\partial(\ell(u,v))_{ps}}\,
\tfrac{\partial(\ell^\wedge(u,v))_{nk}}{\partial(\ell(u,v))_{rr}}\right)
\tfrac{\partial(\ell(u,v))_{ps}}{\partial u}\right)_{|_{(u,v)=(u_j,v_j)}}\,.
\ee
Such two expressions are equal to each other by the reasons pointed
out in the end of the proof of Theorem 2. 
\endproof
\newpage

%
%\vskip 0.2cm
%

\noindent
{\bf Theorem 6.} {\it The variables $(u_j,v_j\equiv y_j)$, $j=1,\ldots,n-1$,
together with the variables $(X,P)$ describing the motion of the 
center-of-mass,
\beq
X:=x_n\,,\qquad P:=\tr\;\ell(u)=\sum_{j=1}^np_j\,,
\label{unvn1}\eeq
constitute the complete canonical set of new (separation) variables.}
\vskip 0.2cm

\noindent
{\bf Proof} repeats the proof of the Theorem 3.
\endproof
\vskip 0.2cm

Consider now nonrelativistic limit of the generating functions $F$ \Ref{FF}
and \Ref{FFF}
in the two simplest cases. In analogy with calculations in the previous Section,
for the case $n=3$ let us define two functions $A_1(u)$ and $A_2(u)$ by the
formulas \Ref{QQ}, or explicitly,
\beq
  A_k(u)=L_{kk}-\frac{L_{3k}\,L_{k,3-k}}{L_{3,3-k}}=p_k+ig\, a_k(u)\,, \qquad
  k=1,2\,,
\eeq
\be
  a_k(u)=\zeta(u)+\zeta(x_3-x_k)+\zeta(x_k-x_{3-k})-\zeta(u+x_3-x_{3-k})\,,
 \qquad k=1,2\,.\ee
The $\pm$-variables are defined by \Ref{xFF}--\Ref{xxFF} and we have the restriction 
\beq
 u_+\equiv x_+\qquad ({\rm mod}\,\Gamma)\,.
\eeq
The generating function $F(y_+,x_+;u_-,x_-)$ (cf. formula (7.12) in \cite{S1})
is then given by the expression
\beq
 F=y_+\,x_++ig\,\log\left[
\frac{\s\left(\frac{x_-+u_-}{2}\right)\s\left(\frac{x_--u_-}{2}\right)
      \s\left(\frac{x_++x_-}{2}\right)\s\left(\frac{x_+-x_-}{2}\right)}%
{\s\left(\frac{x_++u_-}{2}\right)\s\left(\frac{x_+-u_-}{2}\right)
\s(x_-)}\right]\,.
\eeq

Similarly, in the case $n=2$, the normalisation vector is taken as follows:
$$
\vec\a_1=(\,\Phi_u(\xi-x_1),\;\Phi_u(\xi-x_2)\,)\,.
$$
Introduce the functions $A_1(u)$ and $A_2(u)$ by the formulas
\Ref{QQQ}, or explicitly,
\beq
 A_1=L_{11}-\tfrac{\Phi_u(\xi-x_1)}{\Phi_u(\xi-x_2)}\;L_{12}
=p_1+ig\,a_1(u)\,,
\eeq 
\beq
 A_2=L_{22}-\tfrac{\Phi_u(\xi-x_2)}{\Phi_u(\xi-x_1)}\;L_{21}
=p_2+ig\,a_2(u)\,,
\eeq
\be
 a_k(u)=\zeta(u)+\zeta(\xi-x_k)+\zeta(x_k-x_{3-k})-\zeta(u+\xi-x_{3-k})\,,
\qquad k=1,2\,.
\ee
The variables $x_\pm$ are defined in this case as $x_\pm=x_1\pm x_2$
and we have the restriction
\beq
 u_+\equiv x_+-2\xi\qquad ({\rm mod}\,\Gamma)\,.
\eeq
The generating function $F(y_+,x_+;u_-,x_-)$ has the following form
\beq
 F=y_+\,x_++ig\,\log\left[
\frac{\s\left(\frac{x_-+u_-}{2}\right)\s\left(\frac{x_--u_-}{2}\right)
  \s\left(\frac{x_++x_--2\xi}{2}\right)\s\left(\frac{x_+-x_--2\xi}{2}\right)}%
{\s\left(\frac{x_++u_--2\xi}{2}\right)\s\left(\frac{x_+-u_--2\xi}{2}\right)
\s(x_-)}\right]\,.
\eeq
%
%%%%%%%%%%%%%%%%%%%%%%%%%%%%%%  SECTION 7  %%%%%%%%%%%%%%%%%%%%%%%%%%%%%%%%%%%%%%
%
\newsection{Concluding remarks}
\setcounter{equation}{0}
We have performed the separation of variables for the classical
$n$-particle Ruijsenaars system. If we replace the $\s$-function
$\s(x)$ in all the above formulas by $\sin(x)$ ($\sinh(x)$) or
by the identity function: $x\rightarrow x$, then we get 
all the above statements valid for the cases of 
trigonometric (hyperbolic) or rational Ruijsenaars system, 
respectively. 

We have found the explicit generating 
function $F(u|x)$ of the separating canonical transform 
in the cases of two and three particles. It is a challenging 
problem to obtain such a function for $n>3$ in any explicit form.
What is also a problem for possible further studies of this 
integrable system  is to produce a quantum SoV, i.e. to find 
the corresponding kernel ${\cal M}_\hbar(u|x)$ of the quantum 
separating integral operator $M_\hbar$ and related integral 
representation for eigenfunctions of the quantum integrals
of motion $H_j$ (cf. \cite{S1,KS1,KS2,KS3,Factor}). 
%
%%%%%%%%%%%%%%%%%%%%%%%%%%%%%%  SECTION   %%%%%%%%%%%%%%%%%%%%%%%%%%%%%%%%%%%%%%%%
%
\section*{Acknowledgments}
VBK and FWN wish to acknowledge the support of EPSRC.
%
%%%%%%%%%%%%%%%%%%%%%%%%%%%%%%%%%%%%%%%%%%%%%%%%%%%%%%%%%%%%%%%%%%%%%%%%%%%%%%%%%%
%
\bibliographystyle{plain} 

\end{document}